\documentclass[12pt]{article}

\begin{document}


\centerline{\large{\bf MATRIX MODEL THERMODYNAMICS}}

\begin{center}{{\bf Gordon W. Semenoff}\\
 {\it Department of Physics and Astronomy, University of British
Columbia,\\ Vancouver, British Columbia, Canada V6T 1Z1  Email:
semenoff@nbi.dk}}
\end{center}


\begin{center}{Presented at
``Circumnavigating Theoretical Physics'' \\In Memory of Ian Kogan,
Oxford, U.K., January 2004. \\Plenary lecture at ``Quantum Theory
and Symmetries''\\ Cincinnati, Ohio, October 2003 }\end{center}

\textbf{}

\abstract{Some recent work
 on the thermodynamic behavior of the matrix
model of M-theory on a pp-wave background is reviewed. We examine
a weak coupling limit where computations can be done explicitly.
In the large N limit, we find a phase transition between two
distinct phases which resembles a ``confinement-deconfinement''
transition in gauge theory and which we speculate must be related
to a geometric transition in M-theory. We review arguments that
the phase transition is also related to the Hagedorn transition of
little string theory in a certain limit of the 5-brane geometry. }


\section{Prologue}

Ian Kogan was a great friend and I will miss him dearly. Part of
his journey from the Soviet Union to Oxford passed through
Vancouver where he spent a few years. I have great memories of
that time.

Among Ian's very broad range of scientific interests was a
continuing fascination with critical behavior of string theory at
high temperature. One characteristic of strings is that their
density of states increases exponentially at large energies,
\begin{equation}\rho(E)=E^\alpha e^{E/T_H}\end{equation} The
constant $T_H$ is called the Hagedorn temperature. A consequence
of this large density of states is that, depending on the
exponent, $a$, string theory has either an upper limiting
temperature or a phase transition.

Perhaps my favorite of all of Ian's scientific works is an old
result\cite{Kogan:jd}, (found independently in
\cite{Sathiapalan:1986db}) about an interpretation of the Hagedorn
temperature in string theory. In that work  he noted an analogy
between the Hagedorn behavior of strings and the
Kosterlitz-Thouless phase transition for the unbinding of vortices
in the world-sheet sigma model.  He interpreted the latter in
string theory as a disintegration of the worldsheet by
condensation of vortices. Characteristic of Ian's work, this very
original and fascinating idea seemed well ahead of its time,  In
all likelihood, its full import has yet to be realized.

Ian's interest in the high temperature behavior of strings
continued throughout his career. Some of his recent work explored
the use of the AdS/CFT correspondence to understand the phase
structure of string theory at high temperatures
\cite{Abel:1999dy,Abel:1999rq,Barbon:1998cr}.

In this Paper, which I dedicate to Ian, I will  discuss some of my
own recent work on similar topics.

\section{Motivation}

The basic degrees of freedom of string theory and M-theory are
thought to be known and encoded in the BFSS \cite{Banks:1996vh}
matrix model. The model is supersymmetric matrix quantum mechanics
with action
\begin{equation}
S=\int dt{\rm Tr}\left[
\frac{1}{2R}(DX^i)^2+\frac{R}{4\ell_{pl}^6}[X^i,X^j]^2+\bar\psi
D\psi +\frac{R}{\ell_{pl}^3}\bar\psi\Gamma^i[X^i,\psi]\right]
\label{bfss}
\end{equation}
where $i,j=1,...,9$ and all degrees of freedom are $N\times N$
Hermitian matrices.  This is a gauge invariant theory with
covariant time derivative $D=\frac{d}{dt}-i[A,...$. The gauge
theory coupling constant is given in terms of the null
compactification radius $R$ and the eleven dimensional Planck
length $\ell_{pl}$ (or the ten dimensional IIA string coupling
$g_s$ and string length $\ell_s=\sqrt{\alpha'}$) by
\begin{equation}\label{coupling1}
g^2_{YM}=\left( \frac{R}{\ell_{pl}^2}\right)^3 = g_s \ell_s^3
~~,~~\ell_{pl}=g_s^{1/3}\ell_s ~~,~~ R=g_s\ell_s\end{equation} This
model has three uses. It is conjectured to describe a discrete
light-cone quantization of M-theory\cite{Banks:1996vh} where $R$ is
the compactification radius of the light-cone, there are $N$ units of
light-cone momentum and $\ell_p$ is the Planck length of
11-dimensional supergravity. Secondly, and historically a little
earlier\cite{Witten:1995ex}, with parameters suitably re-identified, it
describes the low energy dynamics of a collection of $N$ D0-branes of
type IIA superstring theory.  Finally, it is a matrix regularization
of the light-cone action for the 11-dimensional
supermembrane\cite{Taylor:2001vb}.

One motivation for understanding the behavior of matrix models
such as the BFSS model at finite temperature comes from the
conjecture that their finite temperature states are related to
black hole states of type IIA
supergravity\cite{Klebanov:1997kv,Banks:1997tn}. This idea was
studied in a series of papers by Kabat and
Lowe\cite{Kabat:2000zv}.  They begin with the Beckenstein-Hawking
entropy of a black DO-brane solution of IIA supergravity -- the
area of its even horizon in Planck units.  They convert the
entropy to the free energy, which they then write in terms of
gauge theory parameters to obtain

\begin{equation}
F/T=-4.115 N^2\left( \frac{T^3}{g_{YM}^2N} \right)^{3/5}
\label{entropy}
\end{equation}
A derivation of (\ref{entropy}) from the matrix model would be an
important result, a first principles computation of non-extremal
black hole entropy using string theory. However, one would expect
to find (\ref{entropy}) in a low temperature and therefore strong
coupling limit of the matrix model, making it inaccessible to
perturbation theory.   A variational technique was applied and
claimed approximate agreement with the formula over some range of
temperature\cite{Kabat:2000zv}.

The formula  (\ref{entropy}) is remarkable in three respects.
First, it has the correct dependence on $N$ and the 'tHooft
coupling $g_{YM}^2N$ to be the leading order of the 'tHooft limit
of the gauge theory.  If it could be derived in a perturbative
expansion, it would obtain contributions only from planar Feynman
diagrams. This means that the 'tHooft limit should be part of the
limiting process that would extract the classical physics of black
holes from the matrix model.

Secondly, the scaling with $N$ in (\ref{entropy}) is as if this
0-dimensional gauge theory were in a de-confined phase. This is
particularly true if we assume that we are taking the 'tHooft
limit. The use of the word ``deconfined'' in a theory where there
is no spatial extent over which particles can be separated must be
justified carefully. The gauge theory has a Gauss law constraint
so that quantum states of the Hamiltonian must be singlets under
the gauge symmetry. The gauge field in (\ref{bfss}) enforces this
constraint. The number of singlets at a given energy do not scale
like $N^2$, rather they are of order one.  As an example of this,
consider a matrix harmonic oscillator with Hamiltonian
\begin{equation}\label{Hamiltonian}H=\sum_{i=1}^d\omega{\rm Tr}
(\alpha_i^{\dagger}\alpha_i)\end{equation} and matrix-valued
creation and annihilation operators with algebra
\begin{equation}\label{comm} \left[
\alpha^i_{ab},\alpha^{j\dagger}_{cd}\right]=\delta^{ij}\delta_{ad}
\delta_{bc}\end{equation} States are created by
$\alpha^{i\dagger}_{ab}$ operating on a vacuum $|0>$.  To get a
state with energy $E=n\omega$ we must act with $n$ creation
operators.

The analog of gauge invariance is to require a physical state
condition of invariance under the unitary transformation
$$\alpha^{\dagger}\to u\alpha^{\dagger}u^{\dagger}$$ where $$u\in
U(N)/U(1)$$  We assume that the vacuum state is invariant under
this gauge transform.  Then, physical states are created by
operating with invariant combinations of creation operators. In
the limit $N\to\infty$ all such combinations are traces

\begin{equation} \left[{\rm
Tr}\left(\alpha_i^{\dagger}\right)\right]^{n_1}\left[{\rm
Tr}\left(\alpha_{j_1}^{\dagger}
\alpha_{j_2}^{\dagger}\right)\right]^{n_2}\left[{\rm Tr}\left(
\alpha_{k_1}^{\dagger}\alpha_{k_2}^{\dagger}
\alpha_{k_3}^{\dagger}\right)\right]^{n_3}\ldots|0>\label{state}\end{equation}
where the energy is $$E=\omega(n_1+2n_2+3n_3+\ldots)$$ The number
of these traces with a fixed energy, $E$, does not scale like
$N^2$ as $N\to\infty$, instead it approaches a constant as $N$ is
taken large.  Thus for normal thermodynamic states, one would not
expect the free energy to be of order $N^2$.

However, the number of independent traces does increase rapidly
with $E$. It has been shown\cite{Furuuchi:2003sy} that, in the
large $N$ limit, the oscillator has a Hagedorn-like density of
states at high energy, $$\rho(E)\sim \frac{1}{E}e^{ E/T_H}$$ where
the Hagedorn temperature is $$T_H=\omega/\ln d$$ A similar result
has been found for weakly coupled Yang-Mills
theory\cite{Sundborg:1999ue,Polyakov:2001af,Aharony:2003sx}.

At temperatures higher than $T_H$, the thermodynamic canonical
ensemble does not exist. It could be made to exist by keeping $N$
large but finite. That would cut off the exponential growth in the
asymptotic density of states at some large energy. Then we could
consider a temperature that is greater than $T_H$. Both the energy
and entropy would be dominated by states at and above the cutoff
scale. Then, the divergence of the free energy $\sim N^2$ occurs
as we take the limit $N\to\infty$ at constant temperature (noting
that the Hagedorn temperature does not depend on $N$).

As we shall show in the following, a behavior like this can indeed
by found in the matrix model.  In more conventional terms, it
occurs as a large $N$ Gross-Witten type of phase
transition\cite{Gross:1986iv,Semenoff:1996vm} which is familiar in
unitary matrix models. It is this behavior that we call
``deconfinement''. At this deconfinement transition,
$\lim_{N\to\infty}F/N^2$ jumps from being zero to of order one.

In an adjoint gauge theory such as (\ref{bfss}), there is an order
parameter for confinement, the Polyakov
loop\cite{Polyakov:vu,Grignani:1995hx}.  It is the trace of the
holonomy of the gauge field around the finite temperature
Euclidean time circle,
\begin{equation}
P=\frac{1}{N}{\rm Tr}\left( e^{i\oint A}
\right)
\label{ploop}\end{equation}
This operator gets a nonzero expectation value when a gauge theory is
deconfined.  An interesting question is whether it has a nonzero
expectation value in the BFSS matrix model.  In such a low dimensional
theory, it can only have a nonzero expectation value when $N$ is
infinite.  Indications from weak coupling
computations\cite{Ambjorn:1998zt} are that it has.

The third remarkable fact about (\ref{entropy}) is, though this
formula is thought to apply to the black hole only for a range of
temperatures\cite{Itzhaki:1998dd}, the expression at low
temperature is reminiscent of critical scaling with a critical
temperature $T=0$ and a simple, rational critical exponent.

In spite of the simplicity of these interesting features, there is
no analytic derivation of the formula (\ref{entropy}) from the
matrix model. One of the difficulties in finding a derivation is
the intractability of the model itself. These difficulties are
well-known from previous attempts to analyze its
thermodynamics\cite{Ambjorn:1998zt}.

Before we continue with the matrix model, we comment that, if we
analyze the thermodynamics of M-theory in the rest frame, we would
form the partition function with a Boltzman distribution using the
energy $E=\frac{1}{\sqrt{2}}(P^++P^-)$,
\begin{equation}
Z=\sum_N e^{-N/\sqrt{RT}}e^{-F/\sqrt{2}T}
\label{mtp}\end{equation} where the matrix model free energy is
defined by
\begin{equation} e^{-F/T}={\rm Tr}e^{-H/T}
\end{equation}
In (\ref{mtp}) we have traced over the eigenstates of the
light-cone momentum. This gives the sum over matrix model
partition functions for each $N$ with the exponential factor
$e^{-N/RT}$. The convergence of the sum is clearly dependent on
the nature of the large $N$ limit. If there is a sector of the
matrix model where this limit is like (\ref{entropy}), because of
the negative sign, the sum over $N$ diverges.  It is tempting to
associate this with the non-existence of thermodynamics of a
theory of quantum gravity on asymptotically flat space -- because
of the Jeans instability the space is unstable to collapse to
black holes.  If there were a phase where the free energy did not
become negative and with magnitude growing faster than $N$, it
would be a stable phase.

There are several known behaviors of matrix models in the large N
limit.  For example, there is the
Dijkgraaf-Verlinde-Verlinde\cite{Dijkgraaf:1997vv} limit of matrix
string theory.  That is a strong coupling limit which kills the
off-diagonal degrees of freedom of the matrices. The remaining,
diagonal degrees of freedom are $N$ in number and it can be shown
explicitly\cite{Grignani:2000zm,Grignani:1999sp,Semenoff:2001bi}
that the free energy is negative and is proportional to $N$. Of
course, this is just the correct behavior for a string theory,
there will be a Hagedorn temperature where the large $N$ terms in
the sum in (\ref{mtp}) go from being exponentially suppressed to
growing exponentially.  There are other versions of matrix string
theory\cite{Billo:2001wi,Billo:1999xc,Billo:1999ts,Billo:1999bv}
based on two dimensional Yang-Mills theory where one would expect
a similar behavior.

Another limit where we have a quantitative estimate of the large N
behavior is the 'tHooft limit where N is taken to infinity at the
same time as $g_{YM}$ is taken to zero.  Technically this would be
done by re-defining $\lambda=g_{YM}^2N$ and holding $\lambda$
fixed as when we sum over $N$ in (\ref{mtp}).  Then, the phase
transition that we discuss here is somewhat more violent than the
Hagedorn behavior in string theory.   The matrix part of the free
energy at large $N$ changes from a negative constant to a negative
constant times $N^2$.  The linear in $N$ exponent of the momentum
part doesn't compete with the $N^2$-growth of the matrix model
contribution.

\section{PP-Wave matrix model}

Recently a variant of the matrix model which is conjectured to
describe a discrete light cone quantization of M-theory on a
pp-wave background has been formulated\cite{Berenstein:2002jq}. It
is a 1-parameter deformation of (\ref{bfss}),
\begin{eqnarray}
S=\int dt{\rm Tr}\left[
\frac{1}{2R}(DX^i)^2+\frac{R}{4\ell_{pl}^6}[X^i,X^j]^2+\bar\psi
D\psi
+\frac{R}{\ell_{pl}^3}\bar\psi\Gamma^i[X^i,\psi]\right.\nonumber\\
\left. -\frac{\mu^2}{18R}(X^a)^2-\frac{\mu^2}{72R}(X^{i'})^2
-\frac{\mu}{4}\bar\psi\psi-i\mu\epsilon^{abc}X^aX^bX^c\right]
\label{pp}
\end{eqnarray}
where the indices $a,b,..=1,2,3$ and $i'=4,...,9$. This matrix
model reduces to the BFSS model if we put the parameter $\mu$ to
zero and can be considered a one-parameter deformation of it.  The
main difference between the two is that the action in (\ref{bfss})
has (super)symmetries identical to those of the residual
invariance of 11-dimensional Minkowski space in light-cone
quantization whereas (\ref{pp}) has symmetries appropriate to a
pp-wave spacetime.

The matrix model in (\ref{pp}) has the great advantage that,
unlike (\ref{bfss}), it can be analyzed in perturbation
theory\cite{Kim:2002if}. In (\ref{bfss}), the classical potential
$\sim -{\rm Tr}\left([X^i,X^j]^2\right)$ has flat directions, any
set of matrices which are mutually commuting have zero energy. The
behavior of the degrees of freedom in flat directions must be
understood at the outset of an honest quantum mechanical treatment
of the theory. In (\ref{pp}), these flat directions are removed by
the mass terms. Perturbation theory is accurate in the limit where
the mass gap $\mu$ is large.

The pp-wave space which is a maximally supersymmetric  solution of
11-dimensional supergravity is
$$
ds^2=dx^idx^i-2dx^+dx^--\left( \frac{\mu^2}{9}(x^a)^2 +
\frac{\mu^2}{36}(x^{i'})^2\right)dx^+dx^+
$$
with an additional constant background 4-form flux
$$
F_{+123}=\mu
$$
This space is known to support a spherical membrane solution,
$$
x^+=p^+\tau ~~,x^-={\rm const.}~~,~~
\sqrt{(x^a)^2}=\frac{\ell_{pl}^3}{6}p^+\mu
$$
and a spherical transverse 5-brane
$$
x^+=p^+\tau ~~,~~x^-={\rm const.}~~,~~ (x^{i'})^2=
\ell_{pl}^3\sqrt{\frac{\mu p^+}{6}}
$$

These objects are conjectured to make their appearance in the
solutions of the matrix model.  The membrane is found immediately
in semiclassical quantization.  The classical potential is
(hereafter we set $\ell_{pl}=1$ and measure all dimensional
quantities in Planck units)
\begin{eqnarray}
V=\frac{R}{2}{\rm Tr}\left[ \left(
\frac{\mu}{3R}X^a+i\epsilon^{abc}X^bX^c\right)^2
+\frac{1}{2}\left( i[X^{i'},X^{j'}]\right)^2 +
\right.\nonumber\\
\left.
+\left(i[X^{i'},X^a]\right)^2+
\left(\frac{\mu}{6R}\right)^2(X^{i'})^2
\right]
\end{eqnarray}
It is minimized by
\begin{equation}
X^{i'}_{\rm cl}=0 ~~,~~ X^a_{\rm cl}= \frac{\mu}{3R}J^a
\label{fs}\end{equation} Where $J^a$ is an N-dimensional
representation  of the SU(2) algebra,
$[J^a,J^b]=i\epsilon^{abc}J^c$. In addition, the classical
solution for gauge field must obey the equation
$$
\left[ A_{\rm cl} , J^a_{\rm cl} \right] = 0
$$
If $J^a$ is an irreducible representation of SU(2), by Schur's
Lemma, $A_{\rm cl}=0$.  The gauge symmetry is realized by the
Higgs mechanism. When the representation is reducible, there are
gauge fields which commute with the condensate. This part of
$A_{\rm cl}$ remains undetermined and must still be integrated
over, even to obtain the leading order in the semi-classical
approximation to the partition function.

The configurations in (\ref{fs}) are fuzzy spheres, which are
matrix regularizations of the membranes. The 5-branes on the other
hand do not seem to appear in the perturbative states of the
matrix model.  It has been conjectured\cite{Maldacena:2002rb} that
the 5-branes indeed appear as the large $N$ limit is taken in a
certain way.

First of all, $J^a$ need not be an irreducible representation, but
can contain a number of irreducible components, $$ J^a=\left(
\matrix{ s_1^a &0&0&0 \cr 0&s_2^a&0&0\cr 0&0&s^a_3&0\cr
0&0&0&s_4^a\cr }\right)$$ To get a (multi-)membrane state, the
large $N$ limit is taken by holding the number of representations
fixed and sending the dimension of each of the representations to
infinity.

To get a five-brane state on the other hand, we hold the
dimensions of the representations fixed and repeat them an
infinite number of times to get the large $N$ limit.

An important difference between these limits is in the realization
of the gauge symmetry.  In the classical sectors, the gauge
symmetry is partially realized by the Higgs mechanism, with the
residual symmetry being that which interchanges representations of
the same size. In a membrane state the residual gauge group thus
has finite rank, whereas in a 5-brane state its rank always goes
to infinity. The single 5-brane state is $X^{a}=0$ whereas the
state with $k$ coincident 5-branes has the k-dimensional
representation repeated $N/k \to\infty$ times.  A state with k
non-coincident 5-branes has largest representation k-dimensional
and a number of smaller representations all repeated an infinite
number of times.

The effective coupling constant which governs a semi-classical
expansion about one of the classical ground states is
$$
\lambda= \left( \frac{3R}{\mu}\right)^3 n
$$
where $n$ is the rank of the residual gauge group.  The 5-brane limit
is where we are required to take the weakest coupling limit.  It is
the conjecture of ref.\cite{Maldacena:2002rb} that the
membranes and all other degrees of freedom decouple in this limit and
it isolates the internal dynamics of the 5-brane.

\section{Perturbative expansion}

Now, let us consider a perturbative expansion of the pp-wave
matrix model at finite temperature.  The partition function is the
path integral $$ Z=\int [dA][dX^i][d\psi]e^{-\int_0^\beta d\tau
L[A,X^i,\psi]}$$ where $L$ is the lagrangian with Euclidean time
and $\beta=1/T$ is the inverse temperature.  The bosonic and
fermionic variables have periodic and antiperiodic boundary
conditions, respectively
$$A(\tau+\beta)=A(\tau) ~~,~~X^i(\tau+\beta)=X^i(\tau)
~~,~~\psi(\tau+\beta)=-\psi(\tau) $$ Since the fermions are
antiperiodic, these boundary conditions break supersymmetry.  Of
course this is expected at finite temperature.

We begin by fixing the gauge.  It is most convenient to make the
gauge field static and diagonal,
$$
\frac{d}{d\tau}A_{ab}=0 ~~,~~ A_{ab}=A_a\delta_{ab}
$$
The remaining degrees of freedom of the gauge field are just the
time-independent eigenvalues $A_a$.

The Faddeev-Popov determinant for the first of these gauge fixings
is this gauge fixing is
\begin{equation}
{\det}'\left(-
\frac{d}{d\tau}\left(-\frac{d}{d\tau}+i(A_a-A_b)\right)\right)
\end{equation}
where the boundary conditions are periodic with period $\beta$.
The prime means that the zero mode of time derivative operating on
periodic functions is omitted from the determinant.  The
Faddeev-Popov determinant for diagonalizing the gauge field is the
familiar vandermonde determinant,
$$
\prod_{a\neq b}|A_a-A_b|
$$
Together the two determinants are
$$
\prod_{a\neq b} {\det}'\left(-
\frac{d}{d\tau}\right)\det\left(-\frac{d}{d\tau}+i(A_a-A_b)\right)
$$
where the prime on the determinant indicates that the static mode
is omitted.\footnote{Using zeta-function regularization,
$${\det}'\left(-\frac{d}{d\tau}\right)=\beta$$.}

If we expand about the classical vacuum corresponding to the
single 5-brane, $X_{\rm cl}^a=0=X_{\rm cl}^i$, we find the
partition function in the 1-loop approximation is
\begin{equation}
Z=\int dA_a\prod_{a\neq b} \frac{ {\det}'\left(
-d/d\tau\right){\det}\left(-D_{ab}\right)
 \det^8\left(
-D_{ab}+\frac{\mu}{4}\right) }{ \det^{3/2}\left(
-D_{ab}^2+\frac{\mu^2}{9}\right) \det^{3} \left(
-D_{ab}^2+\frac{\mu^2}{36}\right) }
\end{equation}
where $D_{ab}=\frac{d}{d\tau}-i(A_a-A_b)$.  Using the formula
$$
\det\left(
-\frac{d}{d\tau}+\omega\right)=2\sinh\frac{\beta\omega}{2}
$$
with periodic boundary conditions and
$$
\det \left(
-\frac{d}{d\tau}+\omega\right)=2\cosh\frac{\beta\omega}{2}
$$
with antiperiodic boundary conditions, we can write
\begin{equation}
Z=\frac{1}{N!}\int_{-1/2}^{1/2} d\left(\frac{\beta
A_a}{2\pi}\right)  \prod_{a\neq b} \frac{
[1-e^{i\beta(A_a-A_b)}][1+e^{-\beta\mu/4+i\beta(A_a-A_b)}]^8
 }{ [1-e^{-\beta\mu/3+i\beta(A_a-A_b)}
 ]^3[1-e^{-\beta\mu/6+i\beta(A_a-A_b)}]^6  }
\end{equation}
The factor of $1/N!$ is the volume of the residual discrete gauge
group which permutes the eigenvalues. When $N$ is finite, it might
be possible to do this integral using the method of residues.

However, to apply to the 5-brane, we require the integral when
$N\to\infty$. There are $N$ integration variables $A_a$ and the
action, which is the logarithm of the integrand is generically of
order $N^2$ which is large in the large $N$ limit. For this
reason, the integral can be done by saddle point integration. This
amounts to finding the configuration of the variables $A_a$ which
minimize the effective action:
\begin{eqnarray}
S_{\rm eff}=\sum_{a\neq b} \left( -\ln
[1-e^{i\beta(A_a-A_b)}]-8\ln[1+e^{-\beta\mu/4+i\beta(A_a-A_b)}]
 +\right.\nonumber\\ \left.~~~~+3\ln[1-e^{-\beta\mu/3+i\beta(A_a-A_b)}
 ]+6\ln[1-e^{-\beta\mu/6+i\beta(A_a-A_b)}]\right)
 \label{effectiveaction}
\end{eqnarray}
To study the behavior, it is illuminating to Taylor expand the
logarithms in the phases (this requires some assumptions of
convergence for the first log)
\begin{eqnarray}
S_{\rm eff}= \sum_{n=1}^\infty \frac{
1-8(-)^{n+1}e^{-n\beta\mu/4}-3e^{-n\beta\mu/3}-6e^{-n\beta\mu/6}}{n}|{\rm
Tr}e^{in\beta A}|^2 \label{eff}
\end{eqnarray}
Each term contains the modulus squared of a multiply would
Polyakov loop (\ref{ploop}).  When a coefficient becomes negative,
the loop condenses.  In fact, as we raise the temperature from
zero (and lower $\beta$ from infinity), the first mode to condense
is $n=1$.  This occurs when $$T_C=\frac{\mu}{12}\ln3\approx
.0758533 \mu$$  The condensate breaks a symmetry under changing
the phase of the loop operator.

\section{A closer look at the phase transition}

At temperatures greater than $T_C$ the eigenvalues $A_a$
distribute themselves so that they are clustered near a particular
point on the unit circle.    To examine the possibility, we
consider the equation of motion for the eigenvalues,
\begin{eqnarray}
\omega(e^{0+}z)+\omega(e^{0-}z)+
8\omega(-r^{3}z)+8\omega(-r^{-3}z)= \nonumber\\
=3\omega(r^{4}z) +3\omega(r^{-4}z) +6 \omega(r^{2}z)+6\omega(
r^{-2}z) \label{ge}\end{eqnarray}
 where $r=e^{\beta\mu/12}$ and the resolvent is defined as
 \begin{equation}\label{resolvent}
 \omega(z)=\frac{1}{N}\sum_{a=1}^N \frac{
 z+e^{i\beta A_a}}{z- e^{i\beta A_a}}\end{equation}
  $\omega(z)$ is holomorphic for $z$ away from the unit
circle and has asymptotic behavior, $\omega(\infty)= 1$ and
$\omega(0)= -1$. In the large $N$ limit the poles in $\omega(z)$,
which occur at the location of the elements $e^{i\beta A_a}$,
coalesce to form a cut singularity on a part or perhaps all of the
unit circle. $\omega(z)$ remains holomorphic elsewhere in the
complex plane.

We must remember that equation (\ref{ge}) is valid only when $z$
is one of the gauge field elements $e^{i\beta A_a}$. In that case,
the sum  in (\ref{resolvent}), which turns into an integral in the
large $N$ limit, must be defined as a principal value. In
(\ref{ge}) this is gotten by averaging over approaching the unit
circle from the inside and from the outside.

It is easy to find one exact solution of (\ref{ge}).   If we
consider the case where $e^{i\beta A_a}$ are uniformly distributed
over the unit circle, so that the sum in (\ref{resolvent}) is
symmetric under $z\to e^{i\theta }z$ we can average over the
symmetry orbit to get
\begin{equation}\label{const}
\omega_0(z)
 = \left\{\matrix{ 1 ~~~& |z|>1
 \cr -1 & |z|<1 \cr } \right. \end{equation}
 The result is certianly holomorphic everywhere away from the unit circle
 and is discontinuous on the entire unit circle.

 The resolvent (\ref{const})  is always a solution of (\ref{ge}) for
 any value of r. This is the symmetric, confining solution of the
 matrix model, where the Polyakov loop operator, whose expectation
 value is a particular moment of
$\omega(z)$ for large $z$, vanishes.  We would expect that this
confining solution is only stable if the temperature is low
enough. At some critical temperature it becomes an unstable
solution and there should be other solutions which have lower free
energy.

The confining phase which we discuss above is stable when $r>3$ or
$r<1/3$.  When $r=3$ or $r=1/3$, we can find a 1-parameter family
of solutions,
\begin{equation}\label{critical}
\omega_1(z)=\left\{ \matrix{  -1- az ~~ & ~~~ |z|<1 \cr
 1+a/z & |z|>1 \cr } \right.
 \end{equation}
 This is an acceptable solution when $|a|<2$.\footnote{ Here, $a/2$
 is the expectation value of the Polyakov loop operator which
 must be less than one.}  If we plug it into
 eqn.(\ref{ge}) and assume that $r>1$, we obtain
 \begin{equation}
 (3r^{-4}+8r^{-3}+6r^{-2}-1)(z-1/z)=0
\end{equation}
which is solved when $r=3$.  If we assume $r<1$ we find an
equation which is solved by $r=1/3$.

To examine the phase transition further, we expand about
$r=\infty$.   We expect the transition to occur at $r=3$ which is
not really large, but we will see that corrections are of order
$1/r^4$, at the 1-percent level.

The asymptotic expansions of the resolvent is
\begin{eqnarray}
\omega(z)=1+2\sum_{n=1}^\infty \frac{1}{z^n}\eta_n \\
\omega(z)=-1-2\sum_{n=1}^\infty z^n \eta_{-n}
\end{eqnarray}
where
$$\eta_n \equiv \frac{1}{N}\sum_{a=1}^N e^{in\beta A_a}$$
are the expectation values of the Polyakov loop operator for $n$
windings. If we assume that $r>1$, an asymptotic expansion of the
equation (\ref{ge}) is
\begin{equation}
\omega(e^{0+}z)+\omega(e^{0-}z)=2\sum_{n=1}^\infty\left(
\frac{6}{r^{2n}}+\frac{3}{r^{4n}}+\frac{8}{r^{3n}}\right) \left(
\eta_n z^{-n}-\eta_{-n}z^n\right) \label{approx1}\end{equation}
Remember that this equation is valid only when $z$ is inside the
cut discontinuity of $\omega(z)$ which is assumed to occur on a
segment of the unit circle.

In the large $r$ limit, the right-hand-side of this equation can
be approximated by the leading terms. It is then similar to the
equations for the eigenvalue distributions in adjoint unitary
matrix models which have been solved in the literature
\cite{Semenoff:1996vm,Semenoff:1996ew,Semenoff:1996xg}.

It is easy to find a solution of (\ref{approx1}) if we truncate
the right-hand-side by retaining only the $n=1$ term.  Consider
the semi-circle distribution of Gross and
Witten\cite{Gross:1986iv}\footnote{The spectral density is defined
by
$$
\rho(\theta)=\frac{1}{N}\sum_{a=1}^N \delta(\theta - \beta A_a)
$$
It is normalized so that
$$
\int_{-\pi}^\pi d\theta\rho(\theta)=1
$$ In the large $N$ limit, it becomes a continuous function of
$\theta$ with support on some or all of the interval $[-\pi,\pi]$.
An example is the semicircle distribution, which is given by
$$
\rho_{sc}(\theta)= \left\{\matrix{
\frac{1}{2\pi(1+t)}\cos\frac{\theta}{2}\sqrt{
2(1+t)-4\sin^2\frac{\theta}{2}}& 0\leq \sin\frac{\theta}{2}\leq
\sqrt{\frac{1+t}{2}} \cr 0 &
\sqrt{\frac{1+t}{2}}<\sin\frac{\theta}{2}\leq 1 \cr} \right.
$$
To get (\ref{ss}) we integrate
$$
\omega_{sc}(z)=\int_{-\pi}^\pi d\theta \frac{
z+e^{i\theta}}{z-e^{i\theta}}\rho_{sc}(\theta)
$$
}
\begin{eqnarray}\label{ss}
\omega_{sc}(z)
=\frac{1}{1+t}\left(\frac{1}{z}-z\right)-\frac{1}{(1+t)}\left(
1+\frac{1}{z}\right)\sqrt{1+2tz+z^2}
\end{eqnarray}
This function  has a cut singularity on the unit circle
 between branch points
 $z_\pm = -t\pm i\sqrt{1- t^2}$
where we take $t$ in the range $-1<t<1$.  When $t\to 1$ the
endpoints of the cut touch each other and the cut covers the whole
unit circle. This is the where the Gross-Witten phase transition
occurs in their unitary matrix model.\cite{Gross:1986iv} In their
case, it is a third order phase transition. In the present case it
is a first order phase transition.  The solution that we found
above, when $r=3$, is just a special case of the semicircle
(\ref{ss}) when $t=1$.

Let us explore $\omega_{sc}(z)$ a little more.  First we note that
it obeys $$ \omega_{sc}(1/z)=-\omega_{sc}(z) $$(with the
appropriate change in the sign of the square root). This means
that the $\eta_n=\eta_{-n}=\eta_n^*$ for all $n$.  We can expand
for small $z$,
\begin{equation}
\omega_{sc}(z)=-1-\frac{3-t}{2}z-\frac{(1-t)^2}{2}z^2
-\frac{(1-t)^2(5t+1)}{8}z^3+\ldots
\end{equation}
from which we identify
\begin{equation}
\eta_0=1~,~~\eta_1= \frac{3-t}{4} ~,~~ \eta_2= \frac{(1-t)^2}{4}
~,~~ \eta_3=\frac{(1-t)^2(5t+1)}{16}~...
\label{moments}\end{equation}
 We see that, there is a critical
point at $t=1$.  At that point, $\eta_0=1$, $\eta_{\pm 1}=1/2$ and
$\eta_{|k|>1 }=0$.  This is precisely the value of $t$ for which
the edges of the cut meet, so that the cut covers the entire unit
circle.   This is also precisely the exact solution
(\ref{critical}) which we found when $r=3$, here with the special
value $a=1/2$.

In fact, the semicircle distribution gives a good approximation to
the solution when $r$ is slightly less than 3. For $z$ in the cut,
$$\omega_{sc}(e^{0+}z)+\omega_{sc}(e^{0-}z)
=\frac{2}{1+t}\left(\frac{1}{z}-z\right)$$ If, for the moment, we
truncate the right-hand-side of (\ref{approx1}) to the term with
$n=1$, we see that the equation is solved by the semi-circle
distribution when
\begin{equation}
\frac{4}{(1+t)(3-t)}=\left(8r^{-3}+3r^{-4}+6r^{-2}\right)
\label{approx2}\end{equation}
 Also, remembering that $t$ falls in
the range$-1<t<1$, we get the critical value of $r$, $r_{\rm
crit}=3$ by setting $t=1$.   (\ref{approx2}) has a solution only
when $r<r_{\rm crit}=3$ (and, here we have assumed $r>1$).

When $r=3$, the $n=2$ term on the right-hand-side of
(\ref{approx1}), which we have ignored, contains
$\left(8r^{-3}+3r^{-4}+6r^{-2}\right)=.086$. Thus, we see that, to
an accuracy of about ten percent, the semicircle distribution is
an approximate solution of the model for temperatures just above
the critical temperature.

\subsection{Systematic improvement of the semicircle}

It is clear what has to be done to improve this approximation. We
can begin with an Ansatz for the resolvent which has a single cut
singularity placed on the unit circle
\begin{equation}
\omega(z)=\sum_{n=1}^K \left(a_n\left(z^{-n}-z^n\right)-
b_n\left(z^n-z^{-n-1}\right)\sqrt{ 1+2tz+z^2}\right)
\label{general}\end{equation} To get the general solution, we
should consider all orders by putting $K\to\infty$.  An
approximate solution is found by truncating at some order $K$. We
will see below that this approximate solution is good near the
phase transition. The coefficients in (\ref{general}) must be
arranged so that, in an asymptotic expansion in small $z$,
\begin{enumerate}
\item{}all of the poles of order $1/z^K,...,1/z$ cancel and
$\omega(0)=-1$ so, the asymptotic series then has the form
$$
\omega=-1-2\eta_1 z-2\eta_2 z^2-\ldots
$$
This gives $K+1$ conditions that the $2K+1$ parameters
$(a_n,b_n,t)$ must obey. \item{}From the above expansion, we
determine the moments in terms of the parameters
$$
\eta_1(a_n,b_n,t)~,~~\eta_2(a_n,b_n,t)~,~\ldots~,~~\eta_K(a_n,b_n,t)
$$
\item{}Then we use the equation  (\ref{approx1}), with the
right-hand-side truncated to order $K$ to get $K$ conditions
\begin{eqnarray}
a_1=f(r)\eta_1(a_n,b_n,t)~,~~~~~~~~~~\label{final1}\\
a_2=f(r^2)\eta_2(a_n,b_n,t)~,~~~~~~~~~~\label{final2}
\\...,~~a_K=f(r^K)\eta_K(a_n,b_n,t)
\label{final3}\end{eqnarray} where
$f(r)=\left(\frac{8}{r^3}+\frac{3}{r^4}+\frac{6}{r^2}\right)$.
This gives $K$ further equations which completely determine the
$2K+1$ parameters of the solution.
\end{enumerate}
For example, if we choose $K=2$, requiring that the poles cancel
and $\omega(0)=-1$ yields the three conditions
$$
a_2-b_2=0~,~~ a_1-b_1-tb_2=0~,~~ b_1(1+t)+b_2\frac{1}{2}(1-t^2)=1
$$
We can use these equations to eliminate $b_1,a_1,a_2$
$$
b_1=\frac{t-1}{2}b_2+\frac{1}{1+t}
~~,~~a_1=\frac{3t-1}{2}b_2+\frac{1}{1+t}~~,~~a_2=b_2
$$
 Then, we can calculate the first and second moments by
 considering an asymptotic expansion of $\omega(z)$.  We get
$$
\eta_1=\frac{ (t+1)^3b_2-2(t-3)}{8}
$$
$$
\eta_2=-\frac{b_2}{16}(3t-5)(1+t)^3+\frac{1}{4}(1-t)^2
$$
and finally, using (\ref{final1}) and (\ref{final2}), we get the
equations
$$
\frac{3t-1}{2}b_2+\frac{1}{1+t}=f(r)\left(\frac{
(t+1)^3b_2-2(t-3)}{8}\right)
$$
$$
b_2 =
f(r^2)\left(-\frac{b_2}{16}(3t-5)(1+t)^3+\frac{1}{4}(1-t)^2\right)
$$

Of course, we already know that these equations are solved at the
critical point by $b_2=0,t=1,r=3\to f(r)=1$. If we consider a
value of $r$ somewhat less than the critical value,
$$
f(r)=1+\epsilon
$$
In this case,
$$
f(r^2)=\frac{187}{2187}+\frac{25\epsilon}{162}
$$
We get
$$
t=1-2\sqrt{\epsilon}
$$
and
$$
b_2=\frac{187}{2000}\epsilon
$$

 This demonstrates that, close to the phase
transition, the semicircle distribution gives an accurate
description of the de-confined phase and this description can be
systematically corrected. It would be interesting to explore this
further to determine precise thermodynamic properties of that
phase. The next term in the series on the right-hand-side of
(\ref{approx1}) which we have ignored, since we truncated to order
2, is proportional to $f(r^3)$ with $r\approx 3$, which suggests
that in the vicinity of $r=3$, the error is less than one percent.
However, we caution that this is the case only for $r$ close to
$3$.  When $r=2$, $f(r^3)=.11$.

\subsection{High temperature limit}

 Another limit we could consider
is the high temperature limit where $r\to 1$.  In that case, we
expect that the values of $e^{i\beta A_a}$ are concentrated near a
point. In fact, the case where they are at a single point $$
\omega(z)=\frac{z+\eta}{z-\eta}$$ is a saddle point for all values
of $r$.  However, for $r\neq 1$ it has infinite positive energy,
crossing over to infinite negative energy when $r=1$.  To see that
it is a solution, we note that, in this case the eigenvalue
support is at $z=\eta$ and therefore the variable in (\ref{ge}) is
$\eta$. Then $\omega(e^{0+}\eta)+\omega(e^{0-}\eta)=0$. Also
$\omega(r^s\eta)+\omega(\frac{1}{r^s}\eta)=0$  and (\ref{ge}) is
solved.  It is easy to see from the action that this solution is
unstable for all values of $r$ except $r=1$.

\subsection{Free energy}

 This shows the nature of the phase transition.  At the
critical point, it will turn out that the free energy is
continuous, but the expectation of the Polyakov loop is equal to
$a$ and is ambiguous.  Just below the transition, the theory is
approximately described by the semi-circle distribution for which
the Polyakov loop is 1/4.  So we see that it jumps in value from 0
to 1/4 at the phase transition.  It is for this reason that we
expect the transition to be of first order.  Indeed, by examining
the free energy, we see that it is given by

 \begin{equation}
 \gamma=\frac{1}{N^2}\sum_{a\neq b}\left(-
 \ln(z_a-z_b)-8\ln(z^a-r^3z_b)+3\ln(z_a-r^4z_b)+6\ln(z_a-r^2z_b)\right)
\end{equation}

For the symmetric solution, where $\rho_0(\theta)=\frac{1}{2\pi}$
$$
\gamma_0=0
$$

When $r$ is large, we can expand to get
\begin{equation}
\gamma\approx \frac{1}{N^2}\sum_{a\neq b}\left[-\ln|1-z_a/z_b|-
\left(
\frac{6}{r^2}+\frac{8}{r^3}+\frac{3}{r^4}\right)\frac{z_a}{z_b}+\ldots\right]
\end{equation}
If we keep only the first term in the large $r$ expansion (note
that there is another term which competes with $1/r^4$ which we
ignore for now), this is approximately an adjoint unitary matrix
model of the kind solved in the literature\cite{Semenoff:1996ew}.
It is solved by the semicircle distribution or the symmetric
distribution.  The free energy is
 \begin{equation}
 \gamma\approx \left\{ \matrix{ 0 & r>3 \cr
 \frac{f(r)}{4}\left( 1-\sqrt{1-\frac{1}{f(r)}}\right)-f(r)+
 \frac{3}{4}+\frac{1}{2}\ln\left[f(r)\left( 1+\sqrt{1-\frac{1}{f(r)}}\right)\right]&
 r<3
 \cr }  \right.
 \end{equation}

The value of the Polyakov loop is
\begin{equation}
 \left<\frac{1}{N}{\rm Tr} ~e^{i\oint A}\right>=\left\{ \matrix{ 0 & r>3 \cr
 \frac{1}{2}\left( 1+\sqrt{1-\frac{1}{f(r)}}\right) &
 r<3
 \cr }  \right.
 \end{equation}

\subsection{Symmetry restoration?}

Because of the low dimensionality of the system that we are
discussing, the symmetry breaking which occurs in the de-confined
phase could be destroyed by quantum fluctuations.  In fact, it
would generally be the case in theories with local interactions.

For example, if N is finite, symmetry breaking is not possible.
The phase transition that we have discussed here can only occur
when $N$ is infinite.  Mathematically, we can think of large $N$
as the analog of a large volume limit in a statistical mechanical
system. If $N$ is large but not infinite, the symmetry is not
broken in a mathematical sense but the decay rate of a
non-symmetric state is exponentially suppressed in the volume, in
this case $\sim e^{-N...}$.

The deconfined solution has a spectral density $\rho(\theta)$.
Because of symmetry of the problem under replacing $\theta$ by
$\theta+$constant, there would be a zero mode of the linear
equation for the fluctuations of $\rho(\theta)$, with
wave-function $\psi(\theta)\sim\frac{d}{d\theta}\rho(\theta)$.
This mode would provide the motion which would restore the
symmetry. However, in the case of the semi-circle distribution,
because of the square-root singularity at the edge of the
distribution, this function is not square-integrable, and
therefore not normalizable.

\section{A stack of 5-branes}

 The fluctuation spectrum of the matrices
about a stack of $k$ 5-branes is known. If, rather than the
trivial vacuum, we had chosen the one where the $k$-dimensional
representation of SU(2) is repeated $N/k$ times, the residual
gauge invariance would be $SU(N/k)$. The gauge field would have a
classical solution with $k\times k$ unit matrices times gauge
fields $A_a$ representing the whole blocks. The spectrum is
known\cite{Maldacena:2002rb,Kim:2002zg} and we can again get an
estimate of the Hagedorn temperature\cite{Furuuchi:2003sy}
$$T_H(k=\infty)=.073815...\mu$$  We see that the temperature is
reduced only slightly in this case.  We take this as evidence that
the Hagedorn temperature in this weak coupling limit is
insensitive to th enumber of 5-branes. This seemingly contradicts
the $k$-dependence of the Hagedorn phase transition of little
string theory which has been computed using holography and which
behaves like $T_C\sim \sqrt{k}$.

An explanation for this contradiction can be found in the limit
that we are using.   It is a large N 'tHooft limit and a further
expansion in weak 'tHooft coupling.  In this limit, if we were to
translate the parameters of our model to those that would describe
the NS five-brane in type II string theory, the radius of the
spherical five-brane would be
$$
\frac{r^2}{\alpha'}\sim \lambda^{1/4}
$$
This means that we are expanding about a small, highly curved
five-brane, whereas the usual holographic result for the Hagedorn
temperature is for a large flat 5-brane.  This is a similar
difficulty as the one which appears in the AdS/CFT correspondence
in general.  There, the analog of the matrix model, which is
maximally  supersymmetric Yang-Mills theory, can be readily
analyzed only in the weak coupling limit using an asymptotic
expansion in $\lambda$.  Supergravity and holography give tools
which compute the strong coupling limit, where $\lambda$ is large.
Thermodynamic quantities like the free energy in particular were
computed in both theories and they do not agree with each other
for this reason.

\section{Discussion}

The phase transition in the matrix model is a peculiar one. Closer
analysis reveals that it is of first order.  It is easy to see
from (\ref{ss}) that the expectation value of the Polyakov loop
operator is zero in the symmetric phase but is non-vanishing in
the high temperature phase and approaches a non-zero value there
even as the critical point is approached from above.   However, at
this critical point, there is no co-existence region where one
phase is meta-stable and the other is stable, as is normally the
case for first order phase transitions. To see this, note that the
potential the phase transition occurs where the potential becomes
unstable to perturbations.   Normally, in a first order behavior,
there are two competing vacua, both of which are perturbatively
stable in the transition region and as parameters are varied one
or the other gets lower free energy and is preferred.  Then they
cross there is a phase transition.  In the present case, the
perturbative instability occurs at the same place as the phase
transition.

There is a question as to whether this is an artifact of the
approximation that we have done here, i.e. if we expanded the
effective action for eigenvalues to higher loop order the phase
transition might be more conventional.

Indeed, in other unitary matrix models applied to non-Abelian
Coulomb gases\cite{Semenoff:1996ew,Semenoff:1996xg}, where the
eigenvalues live on a higher dimensional space (a $D=1$ or even
$D>1$ unitary matrix model) there is a coexistence region.

A similar problem afflicts weakly coupled four-dimensional
supersymmetric Yang-Mills theory\cite{Aharony:2003sx}.  In that
case, Witten argued using AdS/CFT that, in the strong coupling
limit of planar Yang-Mills theory, there should be a de-confining
phase transition, identified with the Hawking-Page phase
transition of supergravity on asymptotically anti-de Sitter
space\cite{Witten:1998zw}. Hawking-Page is a normal first order
phase transition where there is the possibility of metastable
phases. At weak coupling, the analysis looks very similar to what
we have done here for the matrix model and the phase transition
has the same nature.  Aharony et.al. have conjectured that the
effect of higher loop corrections (3-loops) to the effective
action in Yang-Mills theory would indeed change the phase
transition to a more conventional one.

Finally, in the matrix model that we have analyzed, there is the
question of whether the phase transition that we have found has
anything to do with collapse to black holes.  The subject of black
holes on pp-wave backgrounds is a murky one which is presently
being sorted out.  In any case, the physics described by classical
gravity should appear at a strong coupling limit, rather than the
weak coupling limit that we have analyzed.  It is tempting to
conjecture that, if there were black holes, collapse to black
holes at finite temperature is what our phase transition would
describe if it persists at strong coupling.

There is the further question of whether our phase transition is
related to the Hagedorn behavior that is seen in the string
spectrum on pp-wave
backgrounds\cite{PandoZayas:2002hh,Greene:2002cd,Sugawara:2002rs,Brower:2002zx,Grignani:2003cs,Hyun:2003ks,Das:2003yq}.
This behavior of course occurs in weakly string theory which is a
limit of the M-theory.  It there were such a relationship, it
would be interesting to ask whether it is related to the formation
of black holes in ten dimensional supergravity near a pp-wave
background\cite{Bigazzi:2003jk,Hashimoto:2004ve}.

Of course of the conjectured relationship with the 5-brane is
valid, then, indirectly, the Hagedorn phase transition of little
string theory is related to horizon formation in the 5-brane
geometry.  In that case, our phase transition and its
thermodynamics indeed describes the black hole, again in a limit
which is far away from previous analysis of such objects.  Note
that we do not analyze the relative stability of membranes and
five-branes.  This question has been addressed in recent
interesting papers\cite{Shin:2004ms,Yee:2003ge}

It is also interesting to ask whether the phase transition that we
have identified is related to recent work which studied phase
transitions for statistical systems of random walks on discrete
groups like the permutation group.\cite{D'Adda:2002ix,
D'Adda:2001nf}

\section*{Acknowledgments}

The work reviewed in this paper was done in collaboration with
Ehud Schreiber and Kazayuki Furuuchi.   I have benefitted from
conversations with Ofer Aharony, Shiraz Minwalla and Mark Van
Raamsdonk. Many of the ideas here derive from an earlier work done
in collaboration with Jan Ambjorn and Yuri Makeenko.


\begin{thebibliography}{0}


\bibitem{Kogan:jd}
Y.~I.~Kogan, ``Vortices On The World Sheet And String's Critical
Dynamics,'' JETP Lett.\  {\bf 45}, 709 (1987) [Pisma Zh.\ Eksp.\
Teor.\ Fiz.\ {\bf 45}, 556 (1987)].

\bibitem{Sathiapalan:1986db}
B.~Sathiapalan, ``Vortices On The String World Sheet And
Constraints On Toral Compactification,'' Phys.\ Rev.\ D {\bf 35},
3277 (1987).




\bibitem{Abel:1999dy}
S.~A.~Abel, J.~L.~F.~Barbon, I.~I.~Kogan and E.~Rabinovici, ``Some
thermodynamical aspects of string theory,'' arXiv:hep-th/9911004.

\bibitem{Abel:1999rq}
S.~A.~Abel, J.~L.~F.~Barbon, I.~I.~Kogan and E.~Rabinovici,
``String thermodynamics in D-brane backgrounds,'' JHEP {\bf 9904},
015 (1999) [arXiv:hep-th/9902058].

\bibitem{Barbon:1998cr}
J.~L.~F.~Barbon, I.~I.~Kogan and E.~Rabinovici, ``On stringy
thresholds in SYM/AdS thermodynamics,'' Nucl.\ Phys.\ B {\bf 544},
104 (1999) [arXiv:hep-th/9809033].


\bibitem{Banks:1996vh}
T.~Banks, W.~Fischler, S.~H.~Shenker and L.~Susskind, ``M theory
as a matrix model: A conjecture,'' Phys.\ Rev.\ D {\bf 55}, 5112
(1997) [arXiv:hep-th/9610043].

\bibitem{Witten:1995ex}
E.~Witten, ``String theory dynamics in various dimensions,''
Nucl.\ Phys.\ B {\bf 443}, 85 (1995) [arXiv:hep-th/9503124].




\bibitem{Taylor:2001vb}For a review, see
W.~Taylor, ``M(atrix) theory: Matrix quantum mechanics as a
fundamental theory,'' Rev.\ Mod.\ Phys.\  {\bf 73}, 419 (2001)
[arXiv:hep-th/0101126].

\bibitem{Klebanov:1997kv}
I.~R.~Klebanov and L.~Susskind, ``Schwarzschild black holes in
various dimensions from matrix theory,'' Phys.\ Lett.\ B {\bf
416}, 62 (1998) [arXiv:hep-th/9709108].


\bibitem{Banks:1997tn}
T.~Banks, W.~Fischler, I.~R.~Klebanov and L.~Susskind,
``Schwarzschild black holes in matrix theory. II,'' JHEP {\bf
9801}, 008 (1998) [arXiv:hep-th/9711005].


\bibitem{Kabat:2000zv}
D.~Kabat, G.~Lifschytz and D.~A.~Lowe,
 ``Black hole thermodynamics from calculations in strongly coupled gauge
theory,'' Phys.\ Rev.\ Lett.\  {\bf 86}, 1426 (2001) [Int.\ J.\
Mod.\ Phys.\ A {\bf 16}, 856 (2001)] [arXiv:hep-th/0007051].


\bibitem{Furuuchi:2003sy}
K.~Furuuchi, E.~Schreiber and G.~W.~Semenoff, ``Five-brane
thermodynamics from the matrix model,'' arXiv:hep-th/0310286.

\bibitem{Sundborg:1999ue}
B.~Sundborg, ``The Hagedorn transition, deconfinement and N = 4
SYM theory,'' Nucl.\ Phys.\ B {\bf 573}, 349 (2000)
[arXiv:hep-th/9908001].


\bibitem{Polyakov:2001af}
A.~M.~Polyakov, ``Gauge fields and space-time,'' Int.\ J.\ Mod.\
Phys.\ A {\bf 17S1}, 119 (2002) [arXiv:hep-th/0110196].


\bibitem{Aharony:2003sx}
O.~Aharony, J.~Marsano, S.~Minwalla, K.~Papadodimas and M.~Van
Raamsdonk,
 ``The Hagedorn / deconfinement phase transition in weakly coupled large N
gauge theories,'' arXiv:hep-th/0310285.


\bibitem{Gross:1986iv}
D.~J.~Gross and E.~Witten, ``Superstring Modifications Of
Einstein's Equations,'' Nucl.\ Phys.\ B {\bf 277}, 1 (1986).

\bibitem{Semenoff:1996vm}For a review, see
G.~W.~Semenoff and R.~J.~Szabo, ``Fermionic Matrix Models,'' Int.\
J.\ Mod.\ Phys.\ A {\bf 12}, 2135 (1997) [arXiv:hep-th/9605140].



\bibitem{Polyakov:vu}
A.~M.~Polyakov, ``Thermal Properties Of Gauge Fields And Quark
Liberation,'' Phys.\ Lett.\ B {\bf 72}, 477 (1978).


\bibitem{Grignani:1995hx}For some examples of computations of the
Polyakov loop operator in other lower dimensional gauge theories,
see: G.~Grignani, G.~W.~Semenoff, P.~Sodano and O.~Tirkkonen,
``Charge Screening and Confinement in the Hot 3D-QED,'' Nucl.\
Phys.\ B {\bf 473}, 143 (1996) [arXiv:hep-th/9512048];
G.~Grignani, G.~W.~Semenoff, P.~Sodano and O.~Tirkkonen, ``Charge
Screening in the Finite Temperature Schwinger Model,'' Int.\ J.\
Mod.\ Phys.\ A {\bf 11}, 4103 (1996) [arXiv:hep-th/9511110];
G.~Grignani, G.~W.~Semenoff and P.~Sodano, ``Confinement -
deconfinement transition in three-dimensional QED,'' Phys.\ Rev.\
D {\bf 53}, 7157 (1996) [arXiv:hep-th/9504105].




\bibitem{Ambjorn:1998zt}
J.~Ambjorn, Y.~M.~Makeenko and G.~W.~Semenoff, ``Thermodynamics of
D0-branes in matrix theory,'' Phys.\ Lett.\ B {\bf 445}, 307
(1999) [arXiv:hep-th/9810170].

\bibitem{Itzhaki:1998dd}
N.~Itzhaki, J.~M.~Maldacena, J.~Sonnenschein and S.~Yankielowicz,
``Supergravity and the large N limit of theories with sixteen
supercharges,'' Phys.\ Rev.\ D {\bf 58}, 046004 (1998)
[arXiv:hep-th/9802042].

\bibitem{Dijkgraaf:1997vv}
R.~Dijkgraaf, E.~Verlinde and H.~Verlinde, ``Matrix string
theory,'' Nucl.\ Phys.\ B {\bf 500}, 43 (1997)
[arXiv:hep-th/9703030].

\bibitem{Grignani:2000zm}
G.~Grignani, P.~Orland, L.~D.~Paniak and G.~W.~Semenoff, ``Matrix
theory interpretation of DLCQ string worldsheets,'' Phys.\ Rev.\
Lett.\  {\bf 85}, 3343 (2000) [arXiv:hep-th/0004194].

\bibitem{Grignani:1999sp}
G.~Grignani and G.~W.~Semenoff, ``Thermodynamic partition function
of matrix superstrings,'' Nucl.\ Phys.\ B {\bf 561}, 243 (1999)
[arXiv:hep-th/9903246].

\bibitem{Semenoff:2001bi}
G.~W.~Semenoff, ``DLCQ strings and branched covers of torii,''
Nucl.\ Phys.\ Proc.\ Suppl.\  {\bf 108}, 99 (2002)
[arXiv:hep-th/0112043].


\bibitem{Billo:2001wi}
M.~Billo, A.~D'Adda and P.~Provero, ``Branched coverings and
interacting matrix strings in two dimensions,'' Nucl.\ Phys.\ B
{\bf 616}, 495 (2001) [arXiv:hep-th/0103242].

\bibitem{Billo:1999xc}
M.~Billo, M.~Caselle, A.~D'adda and P.~Provero, ``Generalized
two-dimensional Yang-Mills theory is a matrix string  theory,''
Nucl.\ Phys.\ Proc.\ Suppl.\  {\bf 88}, 142 (2000)
[arXiv:hep-th/0001076].

\bibitem{Billo:1999ts}
M.~Billo, A.~D'Adda and P.~Provero, ``Matrix strings from
generalized Yang-Mills theory on arbitrary Riemann surfaces,''
Nucl.\ Phys.\ B {\bf 576}, 241 (2000) [arXiv:hep-th/9911249].

\bibitem{Billo:1999bv}
M.~Billo, M.~Caselle, A.~D'Adda and P.~Provero, ``2D Yang-Mills
theory as a matrix string theory,'' arXiv:hep-th/9901053.



\bibitem{Berenstein:2002jq}
D.~Berenstein, J.~M.~Maldacena and H.~Nastase, ``Strings in flat
space and pp waves from N = 4 super Yang Mills,'' JHEP {\bf 0204},
013 (2002) [arXiv:hep-th/0202021].


\bibitem{Kim:2002if}
N.~w.~Kim and J.~Plefka, ``On the spectrum of pp-wave matrix
theory,'' Nucl.\ Phys.\ B {\bf 643}, 31 (2002)
[arXiv:hep-th/0207034].



\bibitem{Maldacena:2002rb}
J.~Maldacena, M.~M.~Sheikh-Jabbari and M.~Van Raamsdonk,
``Transverse fivebranes in matrix theory,'' JHEP {\bf 0301}, 038
(2003) [arXiv:hep-th/0211139].

\bibitem{Kim:2002zg}
N.~Kim and J.~H.~Park, ``Superalgebra for M-theory on a pp-wave,''
Phys.\ Rev.\ D {\bf 66}, 106007 (2002) [arXiv:hep-th/0207061].


\bibitem{Semenoff:1996ew}
G.~W.~Semenoff and K.~Zarembo, ``Adjoint non-Abelian Coulomb gas
at large N,'' Nucl.\ Phys.\ B {\bf 480}, 317 (1996)
[arXiv:hep-th/9606117].

\bibitem{Semenoff:1996xg}
G.~W.~Semenoff, O.~Tirkkonen and K.~Zarembo, ``Exact solution of
the one-dimensional non-Abelian Coulomb gas at  large N,'' Phys.\
Rev.\ Lett.\  {\bf 77}, 2174 (1996) [arXiv:hep-th/9605172].

\bibitem{Witten:1998zw}
E.~Witten,
 ``Anti-de Sitter space, thermal phase transition, and confinement in  gauge
theories,'' Adv.\ Theor.\ Math.\ Phys.\  {\bf 2}, 505 (1998)
[arXiv:hep-th/9803131].

\bibitem{PandoZayas:2002hh}
L.~A.~Pando Zayas and D.~Vaman, ``Strings in RR plane wave
background at finite temperature,'' Phys.\ Rev.\ D {\bf 67},
106006 (2003) [arXiv:hep-th/0208066].

\bibitem{Greene:2002cd}
B.~R.~Greene, K.~Schalm and G.~Shiu, ``On the Hagedorn behaviour
of pp-wave strings and N = 4 SYM theory at finite R-charge
density,'' Nucl.\ Phys.\ B {\bf 652}, 105 (2003)
[arXiv:hep-th/0208163].

\bibitem{Sugawara:2002rs}
Y.~Sugawara, ``Thermal amplitudes in DLCQ superstrings on
pp-waves,'' Nucl.\ Phys.\ B {\bf 650}, 75 (2003)
[arXiv:hep-th/0209145].

\bibitem{Brower:2002zx}
R.~C.~Brower, D.~A.~Lowe and C.~I.~Tan, ``Hagedorn transition for
strings on pp-waves and tori with chemical potentials,'' Nucl.\
Phys.\ B {\bf 652}, 127 (2003) [arXiv:hep-th/0211201].

\bibitem{Grignani:2003cs}
G.~Grignani, M.~Orselli, G.~W.~Semenoff and D.~Trancanelli, ``The
superstring Hagedorn temperature in a pp-wave background,'' JHEP
{\bf 0306}, 006 (2003) [arXiv:hep-th/0301186].

\bibitem{Hyun:2003ks}
S.~j.~Hyun, J.~D.~Park and S.~H.~Yi, ``Thermodynamic behavior of
IIA string theory on a pp-wave,'' JHEP {\bf 0311}, 006 (2003)
[arXiv:hep-th/0304239].


\bibitem{Das:2003yq}
S.~R.~Das, J.~Michelson and A.~D.~Shapere, ``Fuzzy spheres in
pp-wave matrix string theory,'' arXiv:hep-th/0306270.

\bibitem{Bigazzi:2003jk}
F.~Bigazzi and A.~L.~Cotrone, ``On zero-point energy, stability
and Hagedorn behavior of type IIB  strings on pp-waves,'' JHEP
{\bf 0308}, 052 (2003) [arXiv:hep-th/0306102].


\bibitem{Hashimoto:2004ve}
A.~Hashimoto and L.~Pando Zayas, ``Correspondence principle for
black holes in plane waves,'' arXiv:hep-th/0401197.

\bibitem{Shin:2004ms}
H.~j.~Shin and K.~Yoshida, ``Thermodynamics of fuzzy spheres in
pp-wave matrix model,'' arXiv:hep-th/0401014.
\bibitem{Yee:2003ge}
J.~T.~Yee and P.~Yi, ``Instantons of M(atrix) theory in pp-wave
background,'' JHEP {\bf 0302}, 040 (2003) [arXiv:hep-th/0301120].


\bibitem{D'Adda:2002ix}
A.~D'Adda and P.~Provero, ``Two-dimensional gauge theories of the
symmetric group S(n) and branched n-coverings of Riemann surfaces
in the large-n limit,'' Nucl.\ Phys.\ Proc.\ Suppl.\  {\bf 108},
79 (2002) [arXiv:hep-th/0201181].

\bibitem{D'Adda:2001nf}
A.~D'Adda and P.~Provero, ``Two-dimensional gauge theories of the
symmetric group S(n) in the  large-n limit,'' Commun.\ Math.\
Phys.\  {\bf 245}, 1 (2004) [arXiv:hep-th/0110243].


\end{thebibliography}
\end{document}